# On the Characteristic Isolation of Compact Subgroups within Loose Groups of Galaxies

Giovanni C. Baiesi Pillastrini[1]

*Sezione di ricerca Spettroscopia Astronomica – Unione Astrofili Italiani c/o IASF-INAF Via del Fosso del Cavaliere, 100, 00133 Rome, Italy*

[1] *Permanent address: Via Pizzardi, 13 – 40138 Bologna – Italy; Tel. & Fax: +39-051-342351*

*email: gcbp@it.packardbell.org*

Abstract

We have explored the hypothesis that a compact subgroup lying within a dense environment as loose group of galaxies, at a certain stage of its evolutionary history, could be influenced by the action of the tidal field induced by the gravitational potential of the whole system. We argue that the empty ring observed in projection around many compact subgroups of galaxies embedded in larger hosts originates around the spherical surface drawn by the tidal radius where the internal binding force of the compact subgroup balances the external tidal force of the whole system. This effect would tear apart member galaxies situated close to this region determining a marked isolation of the subgroup from the rest of the host group. If so, subsequent evolution of these subgroups should not be affected by external influences as the infall of new surrounding galaxies on them. Following this idea we have developed a statistical method of investigation and performed an application to point out such an effect studying a loose group of galaxies hosting a compact group in its central region. The system UZC 578 / HCG 68 seems to be a fair example of such hypothesized process.

Keywords: galaxies: clusters: individual (HCG 68, UZC 578)

## 1. Introduction

In a recent paper [1] a detailed simulation based on the Millennium Run has been performed in order to investigate the spatial properties of compact groups. It has been found that simulated compact groups with four or more members may arise prevalently as dense configurations within looser galaxy groups and are deeply embedded near the center of their own dark matter halos. The former property has been confirmed by many observational studies [1 and references therein], [2], while the latter states that compact groups sink towards the center of their large dark halos so that, in projection, they appear spatially isolated from their environments. An effect that has been largely observed in previous studies on compact groups. In fact, one of the criteria to identify the famous HCG catalog of 100 compact groups was the 'spatial isolation' which states that no additional neighboring galaxies, in magnitude range or brighter of the group, should appear within a concentric circle of three times the angular diameter of the compact group [3] (note that [1], in order to make a comparison with the observations, selected their sample applying the same criterion). Then, the isolation is a *characteristic* property of those compact subgroups (CS hereafter) embedded in loose galaxy groups (LG hereafter). Such a typical configuration has been confirmed studying another sample of compact groups extracted from the CFA2+SSRS2 redshift survey (RSCGs) where the degree of isolation has been tested *a posteriori* of the selection [4]. We argue that these empty *rings* seen in projection around CSs are the result of dynamical processes driven by tidal fields generated by the gravitational potentials of the host LGs. It is not clear, however, if the CS dark halos could be involved in the process which leads to the observed isolation of CSs from their parent LGs. Many N-body simulations have suggested that the formation, evolution and survival of compact groups are strongly dependent on the environment in which they are embedded. For example, it has been proposed either that compact groups form continually in a single loose group during its collapse and virialization [5] or the replenishment of a compact group should be due to a process of secondary infall of surrounding galaxies [6]. On the other hand, some studies have focused on the role of the dark halo but with different results. For example, has been shown that compact groups are stable configurations if they are embedded in massive dark matter halos [7]. On the contrary, has been suggested a collapsing scenario where, starting from initial conditions as a "maximum expansion", virialization and no primordial common dark halo, the observed properties of the today compact groups are well reproduced [8]. The interesting property found by [1] in their simulations for which the CSs are confined at the center of their dark halos enable us to suppose that the halos may have a role in the formation of the empty rings. However, this hypothesis seems to be contradicted by the fact that the CS scale radius is *uncorrelated* with the virial radius of the dark halo [1] that is, the degree of isolation of the CSs, given by the ratio of the two radii, is independent from the properties of their dark matter halos. Then, we can reasonably suppose that an external force as the *tidal field* generated by the gravitational potential of a LG hosting a CS may have played an important role on the evolutionary history of the CS. This is not a new idea, however, we know from the tidal theory that inside a homogeneous sphere, the radial component of the tidal field induced by the spherical potential is always tensile while in a heterogeneous sphere it may reverse the sign from tensile to compressive if the local density of the core region overcomes the mean density [9]. In principle, at a certain evolutionary phase of a LG, the core region may have had a flat density profile that switched the radial component of the tidal field from tensile to compressive. Then, at this stage, one may hypothesize that member galaxies located in the core region were "compressed" to form a CS within it. However, the nature of compressive tide is transient because the process of matter condensation in the core modifies the density profile from flat to cusp resetting the usual tensile mode. Many studies have explored the consequences of such effect; among others, the effects of the mean tidal field of a cluster of galaxies on the internal dynamics of a disk galaxy traveling through it [10]; the same effect has been investigated for molecular cloud dynamics [11]; the formation of central massive objects in both late-type and early-type galaxies via compressive tide [12] and so on. To study the complete evolutionary history of a CS from its formation to the final stage, i.e. towards the so-called "fossil" groups [13], in the context of the tidal theory (both in compressive and tensile mode), one should take into account the merging effect [14,15] performing high-resolution numerical models, but it is not the aim of this paper. Instead, since gravitational forces are linear, we prefer to search for a statistical approach using optical and kinematic data to analyze the behavior of the tidal field induced by the potential of a LG on a CS through the present-day location of member galaxies as point-tracers of the total mass distribution. We start from the consideration that an existing CS, located approximately in the central region of a host LG, shows by itself that the overall tidal force is insufficient to disrupt it. This means that the internal binding force balances the external tidal field at a certain, say, *zero tidal spherical surface* traced from the CS center by the *tidal radius* where both effects cancel out. Consequently, external member galaxies close to that surface should be torn apart by the radial component of the tide leaving an empty ring around the CS. If this could be the physical process which leads to the characteristic configuration of a CS isolated from to the rest of the host LG, we expect to find the zero tidal surface falling *within* the boundaries of the empty ring. To disentangle this issue, in Ch. 2 we present a statistical method based on the tidal theory in order to study the behavior of the tidal force due to a host potential on a central CS. Besides, in Ch.3 we present a case study as an example of how our hypothesis can be explored. Then, in Ch.4 we discuss our result and inferences drawn regarding the characteristic isolation of a CS due to the action of the external tide. Finally, in Ch.5 the concluding remarks.

## 2. The tidal approximation

The present model assumes that the source of the tidal force is due to a time-independent gravitational potential generated by the "point mass" distribution of external galaxies (i.e. the member galaxies of the host LG not belonging to the CS) centered on the CS. Of course, the CS does not suffer any kind of tidal effect if the LG members have a homogeneous and perfect spherical symmetric distribution. However, this symmetry holds only on average, so that, in principle, there could be a non-vanishing effect due to the deviation from perfect isotropy. Generally, assuming the tidal effect as a static tidal limitation, the tidal radius is defined as the distance from the center of the perturbed object beyond which the tidal effect of the host potential exceed its self-gravity. That is, the surface at which the forces cancel out each other is spatially fixed by the *tidal radius* beyond which the binding force dominates the internal dynamics of the CS, while external galaxies are torn apart by the tidal field. As usually, the Newtonian approximation of the tidal force can be expressed by

$$F_{tidal,a} \equiv -\frac{d^2\Phi_{ext}}{dR_a dR_b} R_b \equiv F_{ab} R_b$$

where $\Phi_{ext}$ is the external potential and $R$ is the radius vector in the CS reference frame. Now, considering the CS at position vector $r_c$ from the observer under the action of $N$ surrounding galaxies of the LG at a position $r_g$ and mass $m_g$, the external tidal potential is given by $\Phi_{ext} = -G\sum_N \frac{m_g}{|r_g - r_c|}$ .

As stated before, the CS is assumed to be a bound structure, so that, the equilibrium $F_{tidal} = F_{binding}$ must be satisfied. Plugging in: $F_{tidal} = \frac{M}{R^2}$ where $M$ is the estimated mass of the CS and $R$ its virial radius. Then, the tidal radius is given by

$$R_t = \left(\frac{M}{F_{tidal}}\right)^{1/2}$$

where $F_{tidal} = |F_{aa} R_a|$ ; $F_{aa}$ are the three eigenvalues corresponding to the principal axes of the 3 x 3 symmetric matrix

$$F_{ab} = \sum_N \left(\frac{m_g}{|r|^3}\right)\left(\frac{3r_a r_b}{|r|^2} - \delta_{ab}\right)$$   where $r = r_g - r_c$ and $\delta_{ab}$ is the Kronecker delta.

Note that a significant underestimation of $F_{tidal}$ can be due to a persisting divergence of its amplitude within the extension of the LG i.e. outside matter may increase (even if little) the net effect of the tidal tensor and/or the effect of selection which can flatten its estimate. In other words, we will obtain, if not the true, the l*ower* $F_{tidal}$ limit as well as (consequently) the *upper* $R_t$ limit! According to this conservative line of analysis, we neglect the contribution to the tidal forces due to a non-zero Lambda cosmology.

Note that if $R/R_t > 1$ such a solution is wrong! The tidal radius cannot be smaller than the virial radius because signs of tidal disruption are not evident! Consequently, one or more sizes used in the computation may be either under or overestimated (e.g. the adopted virial mass and/or radius for the CS; the mass assigned to the external member galaxies of the LG; the redshift distortion, etc.). Instead, if $R/R_t < 1$, this is the necessary condition to demonstrate our hypothesis. However, to be also sufficient, the computed tidal radius should fall *inside* the boundaries of the empty ring. On the contrary, either the estimation of the tidal radius or our hypothesis are wrong.

3. Application

3.1. Our case study

To find support to our hypothesis we have searched for an observable relevant case based on a CS located within a LG. A fair candidate has been identified in the UZC-SSRS2 group catalog [16] noting that one of them, UZC578, shows a very compact subgroup composed of five galaxies (out of 22 belonging to the whole group) located in its central region. Then, using the LEDA database, that object was identified as the well-known compact group HCG68 at $z \cong 0.0078$ cataloged in [3]. A virial mass of $M = 1.48 \times 10^{12} h^{-1} M_{sun}$ and a median projected separation of $0.033 h^{-1} Mpc$ (where $h$ is the Hubble constant in unit of $100 Kms^{-1} Mpc^{-1}$) have been taken from [17]. UZC578 is located at $z \cong 0.0079$ (from Table 3 of the UZC-SSRS2 catalog [16]). It was identified using the friends of friends algorithm with a linking parameter that scales with increasing redshift in order to take into account the galaxy selection function. Of course, to carry on our analysis we should assume that both UZC 578 and HCG 68 are real, bound galaxy systems. This assumption is supported by the fact that they are located within the fiducial limit of $z \leq 0.01$ under which the redshift bias and the selection effect in separation could be considered negligible. However, it should not be taken for granted since, as recently reported [18], an apparently compact group as M60 situated within Virgo cluster turns out to be the result of chance alignments of four galaxies casting many doubts on the reality of compact groups. In any case, HCG68 has been deeply studied [19-24] and its physical consistency has been generally accepted. Besides, HCG68 has been classified as a Type B group, i.e. a system containing a large fraction of interacting and merging galaxies as expected from a real compact group [25]. Finally, the angular separation between the centers of HCG68 and UZC578 is ~10 arcmin showing a slight off-center position. The amplitude of the separation is not so large to exclude the "concentricity" of the two objects since, in principle, galaxies inside a potential well could appear distributed anisotropically with respect to the location from which it originate.

3.2. Simplifying assumptions

We are mainly interested in showing a method of analysis than performing a rigorous computation of the tidal force. Unfortunately, the redshift gives only the radial component of a recession velocity however, an exact evaluation of the tidal force would require the 3D position of the mass tracers. This is the main bias which affects our application because the true positions of member galaxies are not available. To simplify the application we are forced to adopt as a crude distance indicator the projected distances calculated by the conventional $cz \sin\theta / (100h)$ where $cz$ is the heliocentric radial velocity and $\theta$ is the angular separation between the CS centroid and the N*th* surrounding galaxy. All distances have been corrected for the conventional projection factor $\pi/2$. This assumption could be supported by the following considerations: *i)* galaxies within virialized groups having, on average, nearly isotropic orbits [26,27], their peculiar velocities, on average, should not be too large and, *ii)* because of working in a local rest frame where the dominant motion of the objects would be a common flow in the microwave background rest frame, the difference would be only a coordinate translation so that, no corrections are needed for bulk flow effect. Taking into account these approximations, we complete the input data adopting a virial radius $R$ of $0.065 h^{-1} Mpc$ for HCG 68 obtained from the ROSAT X-ray observations [20,28].

3.3. Error in estimating the tidal force

The galaxy mass $m_g$ is the only free parameter. One may assign a common averaged mass to the sample of galaxies under analysis [29], but this approach cannot give information about the uncertainty affecting the evaluation of the tidal force from an unknown mass distribution. Statistical techniques can help us to estimate the error introduced by $m_g$. To assess the variability of the tidal force, $n = 1000$ computations of $F_{tidal}$ have been done assigning to each of the 17 member galaxies of UZC 578 a value of $m_g$ randomly picked from a normally distributed interval $[2 \times 10^{11}, 3 \times 10^{12}] \cdot h^{-1} M_{sun}$ while their projected positions remain fixed. Then, $F_{tidal}$ and its standard deviation $\sigma_{tidal}$ has been computed averaging the $n$ outputs. The mass interval is assumed to approximate the real range of galaxy mass from late to early morphological types excluding extreme values for dwarf and giant objects.

4. Result and discussion

From the computation we have found that the average amplitude of the tidal force $F_{tidal}$ is $3.3 \times 10^{13} \pm 0.7 h Mpc^{-2} M_{Sun}$ and the tidal radius $R_t$ is $0.211 \pm 0.027 h^{-1} Mpc$ (the error is the conventional $1 \sigma_{tidal}$ scatter around the average value). Then, the ratio $R/R_t = 0.3 < 1$. Because the closest external galaxy to the zero-tidal surface defined by $R_t$ is located at a mean distance of $\approx 0.36 h^{-1} Mpc$ from the central structure, this result implicates that the ring around HCG 68 is an empty space without external member galaxies brighter than the magnitude limit. So, the zero-tidal surface cuts the ring into two (almost) equal parts, according to our expectation and supporting, at least in this case, our hypothesis that the isolation of the CSs correlates with the amplitude of the external tidal fields produced by the host LGs.

Even if this analysis could be applied to all contexts where a CS is embedded in a whatever environment, it becomes meaningless if applied to compact groups located in very low density environments. In fact, in such a context, they should appear as very isolated objects surrounding by few, sparse, and far galaxies which cannot yield significant tidal fields.

Nevertheless, many factors may bias the computed value of $F_{tidal}$. As stated before, the tidal force is fairly determined when its cumulative amplitude converges asymptotically within the sampled sphere, i.e. in our case within the boundary of UZC 578. To test it, the values of the tidal force are computed for a set of concentric external spheres of increasing $0.1\, h^{-1} Mpc$ radius from the center of HCG 68. We have found that the development of the cumulative amplitude of the tide as a function of increasing distances from the center of HCG 68 is asymptotically convergent within $\sim 1\, h^{-1} Mpc$ radius indicating that it is well defined and UZC 578 is large enough to incorporate the major share of the gravitational influence. Of course, some bias effects may hide the true convergence, i.e. the asymptote will be approached farther away due to further tidal influence from larger distances. Actually, if UZC 578 itself would be embedded in a larger condensation as a cluster of galaxies unlikely the convergence would be reached within the boundary of the group. Fortunately, this is not the case and we assume that a further, large contribution from the background should be negligible. In any case, the tidal force would result underestimated justifying the above assumption of $F_{tidal}$ as the lower limit where the true tidal perturbation converges to its final value.

Besides, a suspect may arise on the adopted galaxy mass interval used to extract the random values of $m_g$. That is, if $M$ is assumed as the *true* mass for HCG 68, then, a wrong randomization of $m_g$ could alter the amplitude of the tidal force. In other words, to force $R_t \approx R$, the mass interval to randomize $m_g$ must be increased (lower and upper limit) by a factor of $\sim 10$ with respect to that assumed in Ch. 3.3! We reject this possibility because our interval has been selected according to the current galaxy mass estimations obtained from accurate lensing measurements [30-33]. Therefore, we retain unlikely a larger error of $R_t$ than that computed statistically.

5. Concluding remarks

The main accomplishment of this paper is the introduction of a statistical method based on the tidal theory in order to study the hypothesis according to which the evolution of substructures within galaxy groups towards compact objects could be strongly influenced by the tidal field induced by the potential of the whole system. We have demonstrated that, at least in the present case, the characteristic empty ring observed around HCG 68 correlates with the amplitude of the tidal field due to the external mass distribution of the group member galaxies in which it is embedded. In fact, as expected from our hypothesis, the computed tidal radius indicate that the zero tidal surface (where the internal binding force of HCG 68 balances the external tidal field) falls between the inner and outer projected boundaries of the ring. At present, this fact excludes any influences of the external tidal field on the internal evolution of HCG 68. Therefore, what we record today is that the tidal effect induced by group potential seems to be efficient in isolating the central substructure from surrounding member galaxies creating an empty ring depleted of luminous galaxies. However, we cannot obtain information about the former evolutionary phases of the system and, in particular, if the process of compacting of member galaxies located in the central region was influenced by a compressive tidal field. In fact, using kinematic data as point-like mass tracers to account for the mass distribution of the system, we cannot detect any effect of tidal compression within it since, if the tidal radius would approach the reversal point (i.e. the point where a local density may overcome the median density reversing the tidal field from tensile to compressive), it would diverge to infinity and does not exist beyond it [34]. Then, we can only speculate that in a certain evolutionary stage of a galaxy group such a tidal compressive field may be existed. It could be induced by a flat density profile in the core region that switches the tide from tensile to compressive. It may have played a role in compacting member galaxies located in the core, but, as already stated, indemonstrable with our method. At present, our result cannot be generalized since has been derived from a single case and, furthermore, using approximated kinematic data. Then, a definitive confirmation should be obtained with our method testing on a larger and better sample of similar systems using accurate positional data. If it will be confirmed, numerical models predicting a stable configuration for compact groups through infall and continuous replenishment of new surrounding galaxies would be reviewed.

References

[1] McConnachie AW, Ellison SL, Patton DR. Compact groups in theory and practice - I. The spatial properties of compact groups. Mon Not R Astron Soc 2008; 387: 1281

[2] Hickson P. Compact Groups of Galaxies. Annu Rev Astron Astr 1997; 35: 357

[3] Hickson P. Systematic properties of compact groups of galaxies. Astrophys J 1982; 255: 382

[4] Barton E, de Carvalho RR, Geller MJ. Environments of Redshift Survey Compact Groups of Galaxies. Astron J 1998; 116 : 1573

[5] Diaferio A, Geller M, Ramella M. The formation of compact groups of galaxies. I: Optical properties. Astron J 1994; 107: 868

[6] Governato F, Tozzi P, Cavaliere A. Small Groups of Galaxies: A Clue to a Critical Universe. Astrophys J 1996; 458: 18

[7] Athanassoula E, Makino J, Bosma A. Evolution of compact groups of galaxies - I. Merging rates. Mon Not R Astron Soc 1997; 286: 825

[8] Acevez H, Velazquez H. N-Body Simulations of Small Galaxy Groups. Rev Mex Astron Astr 2002; 38: 199

[9] Cowsik R, Ghosh P. Dark matter in the universe - Massive neutrinos revisited. Astrophys J 1987; 317: 26

[10] Valluri M. Compressive tidal heating of a disk galaxy in a rich cluster. Astrophys J 1993; 408: 57

[11] Das M, Jog CJ. Tidally Compressed Gas in Centers of Early-Type and Ultraluminous Galaxies. Astrophys J 1999; 527: 600

[12] Emsellem E, van de Ven G. Formation of Central Massive Objects via Tidal Compression. Astrophys J 2008; 674: 653

[13] Barnes JE. Evolution of compact groups and the formation of elliptical galaxies. Nature 1989; 338: 123

[14] Dekel A, Devor J, Hetzroni G. Galactic halo cusp-core: tidal compression in mergers. Mon Not R Astron Soc 2003; 341: 326

[15] Renaud F, Boily CM, Fleck JJ, Naab T, Theis Ch. Star cluster survival and compressive tides in Antennae-like mergers. Mon Not R Astron Soc 2008; 391: 98

[16] Ramella M, Geller MJ, Pisani A, Da Costa LN. The UZC-SSRS2 Group Catalog. Astron J 2002; 123: 2976. Available from: http://cdsweb.u-strasbg.fr

[17] Hickson P, Mendes de Oliveira C, Huchra JP, Palumbo GGC. Dynamical properties of compact groups of galaxies. Astrophys J 1992; 399: 353

[18] Mamon GA. The nature of the nearest compact group of galaxies from precise distance measurements. Astron Astrophys

2008; 486: 113

[19] Pildis RA, Bregman JN, Schombert JM. Deep Optical Observations of Compact Groups of Galaxies. Astron J 1995; 110: 1498

[20] Pildis RA, Bregman JN, Evrard AE. ROSAT observations of compact groups of galaxies. Astron J 1995; 443: 514

[21] Rampazzo R, Covino S, Trinchieri G, Reduzzi L. Testing the physical reality of binaries and compact groups. Properties of early-type galaxies in groups with diffuse X-ray emission. Astron Astrophys 1998; 330: 423

[22] Mendes de Oliveira C, Bolte M. The Dwarf Galaxy Population of the Hickson Compact Group 68. In: Astronomical Society of the Pacific Conference Series 1999; Vol 176: 122

[23] Osmond JPF, Ponman TJ. The GEMS project: X-ray analysis and statistical properties of the group sample. Mon Not R Astron Soc 2004; 350: 1511

[24] Finoguenov A, Ponman TJ, Osmond JPF, Zimer M. XMM-Newton study of $0.012 < z < 0.024$ groups - I. Overview of the IGM thermodynamics. Mon Not R Astron Soc 2007; 374: 737

[25] Coziol R, Brinks E, Bravo-Alfaro H. The Relation between Galaxy Activity and the Dynamics of Compact Groups of Galaxies. Astron J 2004; 128: 68

[26] Carlberg RG, Yee HKC, Morris SL *et al*. Galaxy Groups at Intermediate Redshift. Astrophys J 2001; 552: 427

[27] Mahdavi A, Geller MJ. A Redshift Survey of Nearby Galaxy Groups: The Shape of the Mass Density Profile. Astrophys J 2004; 607: 202

[28] Mulchaey JS, Davis DS, Mushotzky RF, Burstein D. The Intragroup Medium in Poor Groups of Galaxies. Astrophys J 1996; 456: 80

[29] Raychaudhury S, Lynden-Bell D. Tides, torques and the timing argument. Mon Not R Astron Soc 1989; 240: 195

[30] Coil AL, Newman JA, Cooper MC *et al.* The DEEP2 Galaxy Redshift Survey: Clustering of Galaxies as a Function of Luminosity at $z = 1$. Astrophys J 2006; 644: 671

[31] Guzik J, Seljack U. Virial masses of galactic haloes from galaxy-galaxy lensing: theoretical modelling and application to Sloan Digital Sky Survey data. Mon Not R Astron Soc 2002; 335: 311

[32] Hoekstra H, Hiseh BC, Yee HKC, Lin H, Gladders MD. Virial Masses and the Baryon Fraction in Galaxies. Astrophys J 2005; 635: 73

[33] McKay TA, Sheldon ES, Racusin J *et al.* Galaxy Mass and Luminosity Scaling Laws Determined by Weak Gravitational Lensing. arXiv: 0108013 [astro-ph]

[34] Masi M. On compressive radial tidal forces. Am J Phys 2007; 75: 116